\documentclass[pra,a4paper,twocolumn,floatfix,showpacs]{revtex4}
\usepackage{pstricks,graphicx,amsmath,bbm,mathrsfs,amssymb,times,psfrag}
\newcommand{\ket}[1]{\vert #1 \rangle} \newcommand{\bra}[1]{\langle #1 \vert}

\def\N{\sigma_{+}\sigma_{-}}
\def\L{{\cal L}}
\begin{document}
\title{Experimental estimation of one-parameter qubit gates 
in the presence of phase diffusion}
\author{Davide Brivio, Simone Cialdi, Stefano Vezzoli}
\affiliation{Dipartimento di Fisica dell'Universit\`a degli Studi 
di Milano, I-20133 Milano, Italia}
\affiliation{INFN, Sezione di Milano, I-20133 Milano, Italia}
\author{Berihu Teklu Gebrehiwot, Marco G. ~Genoni, Stefano Olivares}
\affiliation{CNISM UdR Milano Universit\`a, I-20133 Milano, Italia}
\affiliation{Dipartimento di Fisica dell'Universit\`a degli Studi 
di Milano, I-20133 Milano, Italia.}
\author{Matteo G.~A.~Paris}
\affiliation{Dipartimento di Fisica dell'Universit\`a degli Studi 
di Milano, I-20133 Milano, Italia.}
\affiliation{CNISM UdR Milano Universit\`a, I-20133 Milano, Italia.}
\affiliation{ISI Foundation, I-10133 Torino, Italia.}
\date{\today}
%%%%%%%%%%%%%%%%%%%%%%%%%%%%
\begin{abstract}
We address estimation of one-parameter qubit gates in the presence of
phase diffusion. We evaluate the ultimate quantum limits to precision,
seek for optimal probes and measurements, and demonstrate an optimal
estimation scheme for polarization encoded optical qubits.  An adaptive
method to achieve optimal estimation in any working regime is also
analyzed in  details and experimentally implemented.
\end{abstract}
\maketitle
%%%%%%%%%%%%%%%%%%%%%%%%%%%%
\section{Introduction}\label{s:intro}
Suppose you are given a black box which operates on qubits.  The
reconstruction of the corresponding quantum operations
\cite{op1,op2,op3} is critical to verify its actions as a quantum
logic gate \cite{op4} as well as to characterize decoherence processes
\cite{op5}.  Let us consider the case when the action of the device is
described by the unitary $U_\phi =\exp\lbrace - i \sigma_{\boldsymbol
n}\phi \rbrace$ and corresponds to a phase-shift (rotation) $\phi$ about
a known axis ${\boldsymbol n}$. This is the simplest operation on a
qubit and realizes a one-parameter logical gate, which allows, combined
with the Hadamard gate, the transformation of any qubit state into
another.  The quantum characterization
\cite{qtm,gtm,mkq,kw3,hr99,cole05,cole06} of this kind of devices is of
interest for quantum information processing and amounts to estimate the
phase-shift by measuring a suitable observable at the output. In the
following, we fix the reference frame and assume, without loss of
generality, rotations $U_\phi = \exp\lbrace - i \sigma_z\phi \rbrace$
about the $z$ axis.
\par
In ideal conditions the estimation of the phase-shift consists of
preparing a qubit in a known pure state $\varrho =
\ket{\psi_0}\bra{\psi_0}$ and then performing a suitable
measurements on the (pure) shifted state $\varrho_\phi =
U_\phi\varrho\, U_\phi^\dagger$. In a realistic implementation,
however, the propagation of a qubit is unavoidably accompanied by 
some noise, which influences the estimation scheme and usually
degrades the overall precision.  In this paper we address
estimation in the presence of the most detrimental kind of noise
for a phase gate,
i.e. non dissipative phase noise, which destroys the off diagonal
elements of the density matrix and thus the information on the
imposed phase-shift.  We evaluate the ultimate quantum limits to
precision in the presence of noise and determine both the optimal
preparation of the probe qubit and the optimal measurement to be
performed at the output.  The optimal estimation scheme is then
experimentally demonstrated for polarization encoded 
 optical qubits, together with
an adaptive method to achieve optimal estimation in any working
regime.
\par
The paper is structured as follows. In section \ref{s:QPE} we
describe the system under investigation focusing on the ultimate
limits to precision of phase-shift estimation in the presence of
phase noise.  The quantum Cram\'er-Rao limit as well as the
optimal quantum estimator are explicitly given. Section
\ref{s:spin} introduces a realistic scenario for qubit
phase-shift estimation: an estimation scheme based on spin
measurements is described in detail and the corresponding Fisher
information is derived. In section \ref{s:SimExp} we turn our
attention to optical qubit systems and describe phase-shift
estimation under two different approaches: the inversion method
and the Bayesian estimation.  We also investigate numerically the
robustness of the Bayesian analysis w.r.t. the inversion method
in the non asymptotic regime, i.e., in the case of small amounts
of data. Section \ref{s:Exp} is devoted to the experimental
demonstration of the optimal scheme for phase-shift estimation,
and of the adaptive method to achieve the quantum Cram\'er-Rao
bound in any working regime.  Section \ref{s:concl} closes the paper 
with some concluding remarks.
%%%%%%%%%
\section{Phase-shift estimation in qubit systems}\label{s:QPE}
\begin{figure}[h!]
\includegraphics[width=0.99\columnwidth]{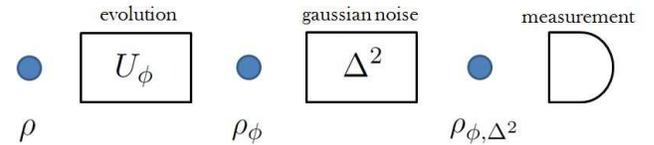}
\caption{Schematic diagram of the measurment scheme.}
\label{f:sketch}
\end{figure}

The measurement scheme we are going to address is schematically
depicted in Fig. \ref{f:sketch}. 
\par
One has a single qubit, initially 
prepared in the pure state $\varrho = \ket{\psi_0}\bra{\psi_0}$, 
that undergoes an unknown phase-shift $\phi$ imposed by the 
unitary $U_\phi$. Before being measured the shifted state $\varrho_\phi =
U_\phi\varrho\, U_\phi^\dagger$ is degraded
by a non dissipative phase noise occurring during the propagation. 
The effect of this kind of noise on the qubit density matrix, given
a noise factor $\gamma$, can be 
described by the following Master equation (ME):
\begin{equation}\label{ME} 
\dot \varrho_{\phi,\Delta^2} = \gamma\, \L[\N]\varrho_{\phi,\Delta^2},
\end{equation}
where $\L[A]\varrho_{\phi,\Delta^2} = \frac12 \left\{ [A\varrho_{\phi,\Delta^2}, 
A^\dag] +
[A, \varrho_{\phi,\Delta^2} A^\dag] \right\}$, and $\Delta^2=\gamma t /2$ 
is, as we will see in the following lines, the effective noise factor.
Since $\L[\N]$ and $\sigma_z$ commute, we can focus the on the evolution of
$\varrho$, i.e., $\dot \varrho_{\Delta^2} = \gamma\, \L[\N]\varrho_{\Delta^2}$.
Upon writing $\varrho_{\Delta^2}$ in the eigenbasis of $\sigma_z$, the ME leads 
to differential equations for the matrix elements 
$\varrho_{nm}(t) = \bra{n} \varrho_{\Delta^2} \ket{m}$, where
$\dot\varrho_{nm}(t) = -\frac12\, \gamma\: (n-m)^2 \varrho_{nm}(t)$
whose solutions read:
\begin{equation}\label{g:sol}
\varrho_{nm}(t) = e^{- \Delta^2(n-m)^2}\varrho_{nm}(0).
\end{equation}
where $\varrho_{nm}(0)$ are the initial density matrix elements.
From Eq.~(\ref{g:sol}) it is clear that, whereas the diagonal elements 
are left unchanged by the evolution under ME (\ref{ME}), and, in turn,
energy is conserved, the off-diagonal ones are progressively destroyed. 
Finally, the solution of Eq.~(\ref{ME}) is $\varrho_{\phi,\Delta^2} = U_\phi 
\varrho_{\Delta^2} \, U_\phi^{\dag}$. Since we can consider the noise factor
$\Delta^2$ as a fixed parameter, in the following we will not write 
explicitely the dependence on it. 
\par
It is worth noting that the same evolution as (\ref{g:sol}) can be
also obtained by the application of a random, zero-mean
Gaussian-distributed phase-shift to a quantum state.  Since the phase
shift of an amount $\varphi$ is described by the unitary operator
$U_\varphi \equiv \exp(-i \varphi\, \sigma_z)$, we can write the state
degraded by the Gaussian phase noise as follows:
\begin{align}
\varrho_{\rm Gn} &= \int_{\mathbbm R}\!\!\! d\varphi\,
\frac{e^{-\varphi^2/(4 \Delta^2)}}{\sqrt{4 \pi \Delta^2}}\,
U_\varphi \varrho(0) U_\varphi^{\dag} \label{g:noise}\\
&= \sum_{nm}\int_{\mathbbm R}\!\!\! d\varphi\,
\frac{e^{-\varphi^2/(4 \Delta^2)}}{\sqrt{4 \pi \Delta^2}}
e^{-i\varphi (n-m)}\, \varrho_{nm}(0) \ket{n}\bra{m}\\
&= \sum_{nm} e^{-\Delta^2 (n-m)^2}\,\varrho_{nm}(0) \ket{n}\bra{m},
\end{align}
which is the same as in Eq.~(\ref{g:sol}). This point will be useful
for the experimental demonstration.
\par
The goal of an estimation problem is not only retrieve the actual
value of the unknown parameter (the phase shift in our case), but
obtain this information with the minimum uncertainty. 
The ultimate limit to the precision one can reach, i.e., the minimum
variance, is given by the quantum Cram\'er-Rao bound
\cite{CR:1,CR:2,CR:3,CR:4}:
\begin{equation}
\hbox{Var}[\phi] = [N\,H(\phi)]^{-1},
\end{equation}
where $N$ is the number of measurements and
$H(\phi)$ is the quantum Fisher information (QFI). 
If we choose
a pure probe state, $\varrho = \ket{\psi_0}\bra{\psi_0}$, then
the QFI equals four times the fluctuation of $\sigma_z$ i.e
one simply has $H = 4 (1- \bra{\psi_0} \sigma_z\ket{\psi_0}^2)$. More in
general, for mixed states, $H(\phi)$ can be written as
\cite{par:QEQT:08}:
\begin{equation}\label{q:fisher}
H(\phi) = 2 \sum_{n\ne m}
\frac{(\lambda_n - \lambda_m)^2}{\lambda_n + \lambda_m}
|\langle \psi_m (\phi)\vert \partial_\phi \psi_n (\phi)\rangle|^2,
\end{equation}
where the $|\psi_n (\phi)\rangle$'s are the eigenvectors of the 
state $\varrho_\phi = U_\phi\varrho\, U_\phi^\dag$ and $\lambda_n$
the correspoding eigenvalues.
\par
The spectral decomposition of $\varrho_\phi$ reads as follows
\begin{equation}
\varrho_\phi = \sum_n \lambda_n \ket{\psi_n(\phi)}\bra{\psi_n(\phi)} =
\sum_n \lambda_n U_\phi\ket{\psi_n}\bra{\psi_n}U_\phi^{\dag}\,.
\end{equation}
$|\psi_n \rangle$ being the eigenvectors of the initial state.
If we decompose $\ket{\psi_n(\phi)}$ in the standard basis as follows:
\begin{equation}\label{ev:dec}
\ket{\psi_n(\phi)} = U_\phi \ket{\psi_n}= U_\phi\sum_k r_{nk} \ket{k},
\end{equation}
then, by substituting (\ref{ev:dec}) into the eigenvalues equation
$\varrho_\phi \ket{\psi_n(\phi)} = \lambda_n \ket{\psi_n(\phi)}$,
after some algebra we obtain:
\begin{equation}
\sum_k \varrho_{nk}(0)\, e^{-\Delta^2(n-k)^2}r_{qk} = \lambda_q r_{qn}
\quad \forall n.
\end{equation}
Moreover, since $\ket{\partial_\phi \psi_n(\phi)} =
i \sum_k k\, r_{nk}\, e^{i k \phi} \ket{k}$, we have [see Eq.~(\ref{q:fisher})]:
\begin{equation}
|\langle \psi_m(\phi) \vert \partial_\phi \psi_n(\phi) \rangle|^2 =
\left|\sum_k k\, r_{mk}\, r_{nk}\right|^2.
\end{equation}
Finally, given $\lambda_n$ and $r_{nk}$, we can evaluate the QFI, which
is, as expected for a shift parameter, 
independent of $\phi$.
\par
Generally, the calculation of the eigenvalues $\lambda_n$ and the
coefficients $r_{nk}$ of the standard basis decomposition
(\ref{ev:dec}) is a difficult task, which can be performed by means
of numerical analysis. However, in the case of qubit systems,
all the calculations can be carried out analytically, as we are
going to show in the following.
\begin{figure}[h!]
\includegraphics[width=0.75\columnwidth]{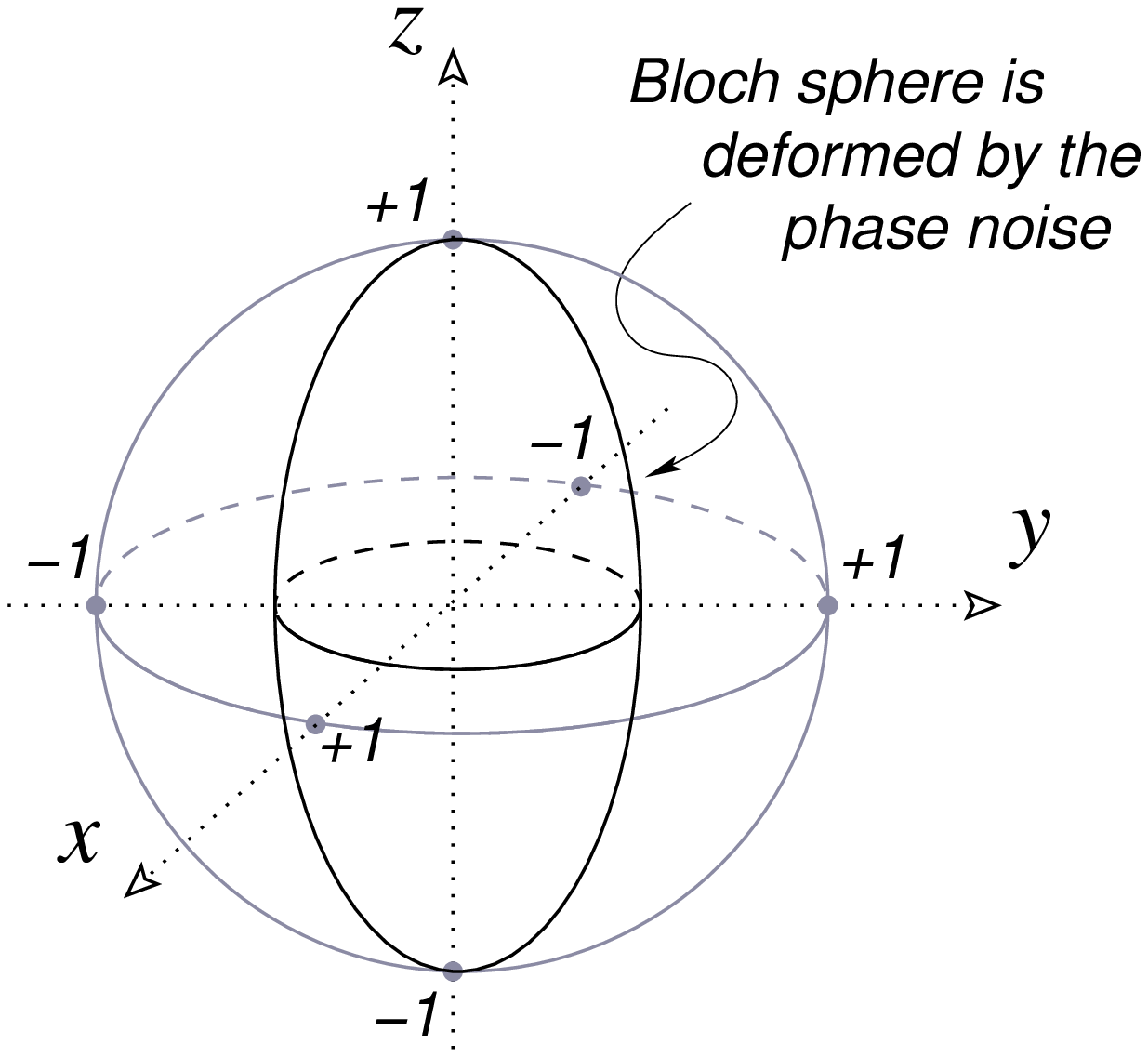}
\caption{\label{f:QBnoise} Effect of the phase destroying ME on the
Bloch sphere: the $z$ component is left unchanged while $x$ and $y$ ones
are scaled by a factor $e^{-\Delta^2}$. See the text for details.}
\end{figure}
\par
Upon writing the initial qubit state in the Bloch sphere representation
\begin{equation}
\varrho = \frac{{\mathbbm 1} + {\boldsymbol r}\cdot {\boldsymbol \sigma}}{2},
\end{equation}
where ${\mathbbm 1}$ is the $2\times 2$ identity matrix,
${\boldsymbol r} = (r_x,r_y,r_z)$, $|{\boldsymbol r}|^2 \le 1$, and
${\boldsymbol \sigma} = (\sigma_x,\sigma_y,\sigma_z)$ is the
Pauli matrices vector, then its evolution under the action of the
ME (\ref{ME}) can be reduced to the following transformation of the
Bloch vector ${\boldsymbol r}$ \cite{NC}:
\begin{equation}\label{bloch:v}
(r_x,r_y,r_z) \to
(r_x e^{-\Delta^2},r_y e^{-\Delta^2},r_z),
\end{equation}
i.e., the Bloch sphere is deformed is such a way that 
the $z$ component is left unchanged while $x$ and $y$ ones
are scaled by a factor $e^{-\Delta^2}$ (see Fig.~\ref{f:QBnoise}).
Now, due to symmetry considerations, without lack of generality we
will focus our attention on the pure state with Bloch vector:
\begin{equation}
{\boldsymbol r} = (\sin 2\theta, 0,\cos 2\theta),
\end{equation}
$2\theta$ being the azimuthal angle ($\theta = 0$ and $\theta = \pi/2$
correspond to the north and south poles of the Bloch sphere,
respectively).
\par
In the density matrix representation (choosing the $\sigma_z$ eigenvectors
basis), after the phase-noise evolution, we have:
\begin{equation}\label{q:ev}
\varrho = \left(
\begin{array}{cc}
\cos^2 \theta & e^{-\Delta^2} \cos \theta \sin \theta \\
e^{-\Delta^2} \cos \theta \sin \theta & \sin^2 \theta
\end{array}
\right).
\end{equation}
The two eigenvalues are:
\begin{equation}\label{l:pm}
\lambda_{\pm} = \frac12 \left(
1\pm\frac{1+f(\theta,\Delta^2)}{\sqrt{2}}
\right),
\end{equation}
with $f(\theta,\Delta^2)=\sqrt{e^{-2\Delta^2}+(1-e^{-2\Delta^2})\cos 4\theta}$,
and the corresponding eigenvectors read:
\begin{align}
\ket{\psi_{\pm}} &= 
\frac{1}{Z_{\pm}} \left(
\begin{array}{c}
g_{\pm}(\theta,\Delta^2)\\
1
\end{array}
\right), \\
&= \frac{1}{Z_{\pm}} \big[
g_{\pm}(\theta,\Delta^2) \ket{+1/2}+\ket{-1/2}
\big],
\end{align}
$\langle \psi_{+} \vert \psi_{-} \rangle = 0$, and:
\begin{subequations}\label{g:pm}
\begin{align}
g_{\pm}(\theta,\Delta^2) &= \cos 2\theta \pm
f(\theta,\Delta^2)/(\sqrt{2}\sin 2\theta)\\
Z_{\pm} &= \sqrt{1+[g_{\pm}(\theta,\Delta^2)]^2}
\end{align}
\end{subequations}
Substituting the previous equations into Eq.~(\ref{q:fisher})
we obtain (remember that for qubit systems $k=\pm 1/2$):
\begin{equation}\label{QFI:qubit}
H(\theta,\Delta^2) = e^{-2\Delta^2} \sin^2 2\theta\,,
\end{equation}
which reaches the maximum for $\theta = \pi/4$: the best states for
phase estimation also in the presence of phase noise are the equatorial
ones, i.e., the states laying in $x$--$y$ plane of the Bloch sphere.
\begin{figure}[h!]
\includegraphics[width=0.8\columnwidth]{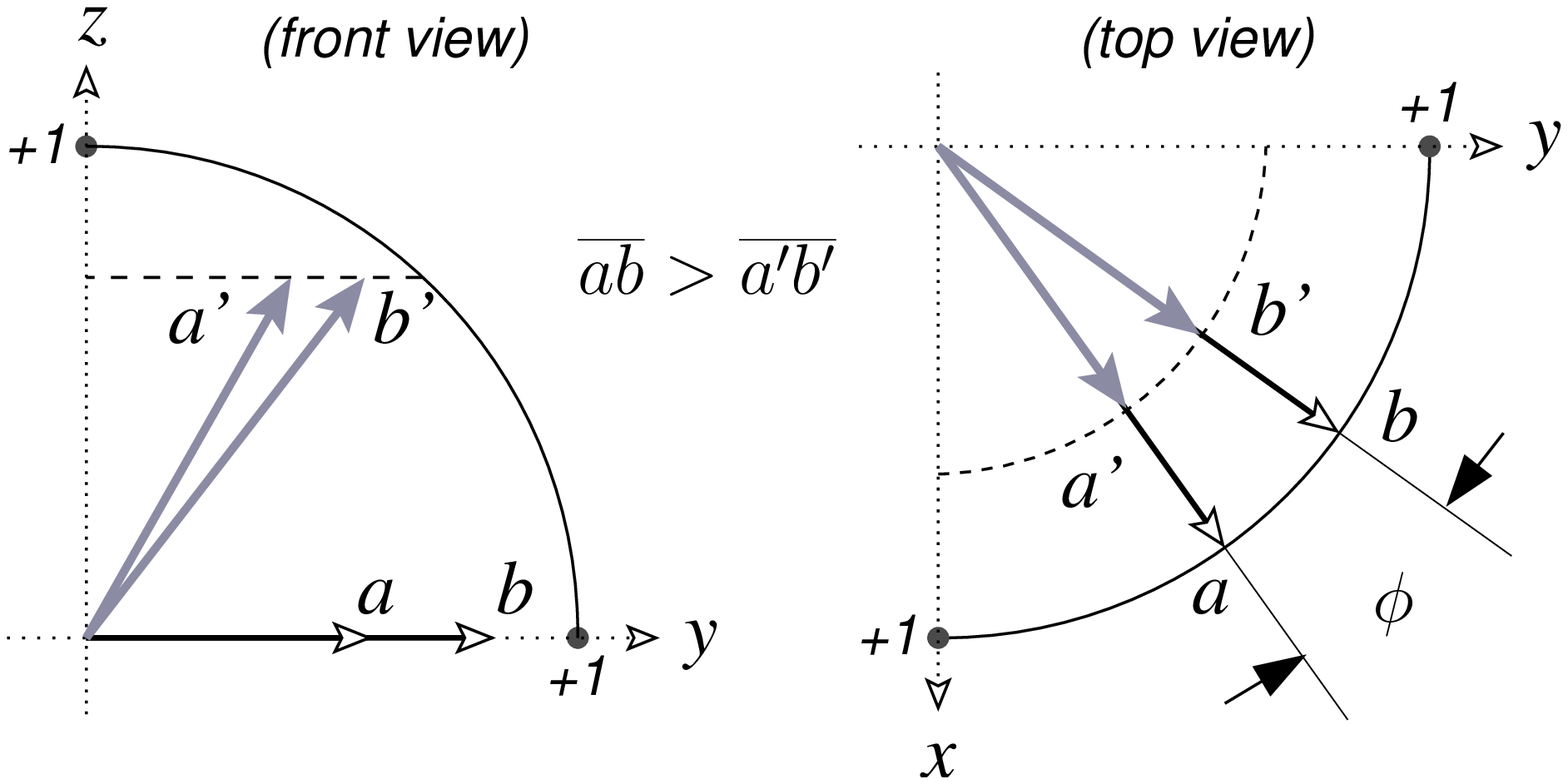}
\caption{\label{f:EQvsS} The distance $\overline{ab}$ on the Bloch
sphere between two equatorial states separated by a phase shift $\phi$
is larger than the distance $\overline{a' b'}$ between non equatorial
states separated by the same amount $\phi$. The same conclusion holds
in the presence of phase noise: best estimation is achieved involving
equatorial states.}
\end{figure}\par
Since the Bures metrics, and then, the Bures distance 
\cite{bures:1,bures:2,bures:3,bures:4,bures:5,bures:6,bures:7} between
states, is proportional to the QFI \cite{somm:03,par:QEQT:08}, this 
result can be easily understood under a geometrical point
of view. If we choose two states, one equatorial and the other not, and shift
them by the same amount $\phi$, then the {\em distance} on the Bloch sphere
between the initial states and the shifted counterparts is larger
for the equatorial states (see Fig.~\ref{f:EQvsS}) which, n turn, allow 
better estimation.
\par
The optimal quantum estimator leading to (\ref{QFI:qubit}) can be written
as \cite{par:QEQT:08}:
\begin{equation}
O_{\phi} = \phi {\mathbbm I} + \frac{L_\phi}{H(\theta,\Delta^2)},
\end{equation}
where we introduced the symmetric logarithmic derivative:
\begin{equation}
\partial_\phi \varrho_\phi =
\frac{L_\phi \varrho_\phi + \varrho_\phi L_\phi}{2},
\end{equation}
and $\varrho_\phi$ is the state of the qubit
initially prepared in a pure state with Bloch vector $\boldsymbol r$
given by (\ref{bloch:v}), shifted by the application of the unitary
transformation $U_\phi = \exp(-i \phi \sigma_{+}\sigma_{-}) =
\exp[-\frac{i}{2} \phi (\sigma_z + {\mathbbm I})]$  and
degraded by the evolution through the noisy environment:
\begin{align}\label{gen:q}
\varrho_\phi
= \left(
\begin{array}{cc}
\cos^2\theta & e^{-i \phi-\Delta^2} \cos \theta \sin\, \theta \\
e^{i \phi-\Delta^2} \cos \theta \sin\, \theta & \sin^2\theta
\end{array}
\right).
\end{align}
One finds:
\begin{equation}\label{gen:L}
L_\phi = i\,
\frac{2 g_{+} g_{-} (g_{+} - g_{-}) (\lambda_{-} - \lambda_{+}) }
{Z_{+}^2 Z_{-}^2 (\lambda_{+} + \lambda_{-}) }
\left(
\sigma_{+} e^{i\phi} - \sigma_{-} e^{-i\phi}
\right),
\end{equation}
where $\lambda_\pm$, $g_{\pm} \equiv g_{\pm}(\theta,\Delta^2)$ and
$Z_{\pm}$ are given in (\ref{l:pm}) and (\ref{g:pm}), respectively.
If we choose $\theta = \pi/4$, Eq.~(\ref{gen:L}) reduces to:
\begin{equation}
L_\phi = i\,  e^{-\Delta^2}
\left(
\sigma_{+} e^{i\phi} - \sigma_{-} e^{-i\phi}
\right).
\end{equation}
%%%%%%%%%%%%
\section{Phase-shift estimation by spin measurement}\label{s:spin}
\begin{figure}[h!]
\includegraphics[width=0.75\columnwidth]{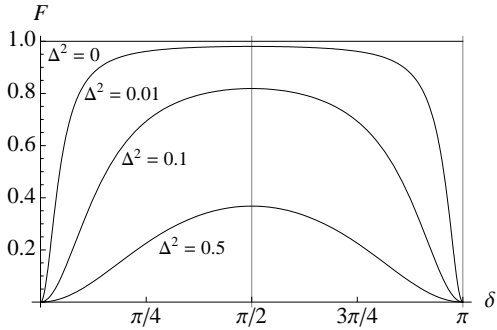}
\caption{\label{f:FQ} Plot of the Fisher information $F(\phi,\Delta^2)$
in the case of equatorial qubit probe states
($\theta = \pi/4$) as a function of $\delta = \alpha - \phi$ and
different values of $\Delta^2$.}
\end{figure}

Let us now consider a realistic scenario where, in order to estimate
$\phi$, we measure the spin in a generic direction in the plane,
i.e. the observable 
\begin{equation}\label{meas:obs}
\Theta_\alpha = \sigma_x\cos\alpha + \sigma_y \sin\alpha,
\end{equation}
whose eigenvectors $\ket{\Sigma_{\pm}(\alpha)}$, 
$\Theta_\alpha \ket{\Sigma_{\pm}(\alpha)} = \pm \ket{\Sigma_{\pm}(\alpha)}$,
are:
\begin{equation}
\ket{\Sigma_{\pm}(\alpha)} = \frac{1}{\sqrt{2}}
\left( e^{-i\alpha} \ket{+1/2} \pm \ket{-1/2} \right).
\end{equation}
For the qubit of Eq.~(\ref{gen:q}), the probabilities to obtain the
outcomes $\pm1$ given the phase shift $\phi$ read:
\begin{align}
P_{\Delta^2}(\pm1 | \phi) &=
\hbox{Tr}[ \ket{\Sigma_{\pm}(\alpha)} \bra{\Sigma_{\pm}(\alpha)}\,
\varrho_\phi]\\
&= \frac{1}{2} \left[ 1\pm e^{-\Delta^2} \cos (\alpha-\phi) \sin 2\theta\right],
\label{prob:pm}
\end{align}
and the expectation value is:
\begin{align}
\langle \Theta_\alpha \rangle &=
\hbox{Tr}[ \Theta_\alpha \, \varrho_\phi) ]
= e^{-\Delta^2} \cos(\alpha -\phi) \sin 2\theta.
\end{align}
The corresponding Fisher information turns out to be:
\begin{align}
F(\phi,\Delta^2) &= \sum_{k=\pm1} P_{\Delta^2}(k | \phi)
\left[\partial_\phi \ln P_{\Delta^2}(k | \phi) \right]^2 \\
&= \frac{e^{-2\Delta^2} \sin^2(\alpha-\phi) \sin^22\theta}
{1- e^{-2\Delta^2} \cos^2(\alpha-\phi) \sin^2 2\theta},
\label{fish:qb}
\end{align}
which is plotted in Fig.~\ref{f:FQ} in the case of equatorial probe states
($\theta = \pi/4$) as a function of $\delta = \alpha - \phi$ and
different values of $\Delta^2$.
\par
\begin{figure}[h!]
\includegraphics[width=0.65\columnwidth]{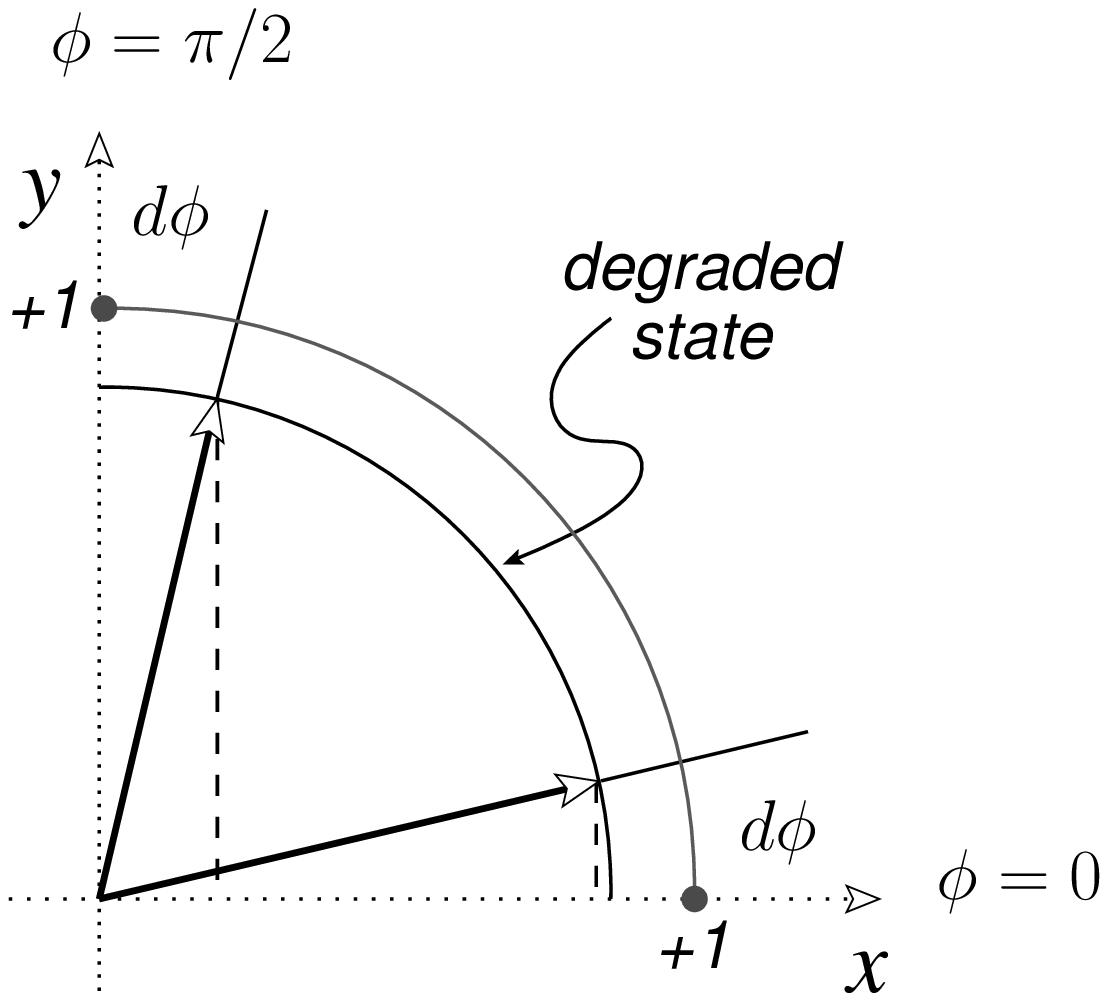}
\caption{\label{f:sigmaX} Sensitivity of $\sigma_x$ measurement to
a small change $d\phi$ in the phase-shift. The projections
onto $x$-axis are the expectations $\langle \Theta_0 \rangle$, with
$\Theta_0 = \sigma_x$. For a fixed change $d\phi$, the change of
$\langle \Theta_0 \rangle$ at $\phi=0$ is smaller than the one at
$\phi = \pi/2$. For the sake of simplicity we sketch only the upper right
quarter of the Bloch sphere $x$-$y$ equatorial plane.}
\end{figure}

As we can see, in the absence of noise ($\Delta^2 = 0$) one has $F=1$,
$\forall \alpha,\phi$, i.e., the Fisher information is equal to the
QFI $H$ in Eq.~(\ref{QFI:qubit}) \cite{yurke86,pezze07}. 
When noise affects the propagation,
the maximum of $F$, that corresponds to the QFI $H$, is achieved for
$\delta = \alpha-\phi = \pi/2$, while goes to zero as $\alpha-\phi =
k\, \pi$, $k\in{\mathbbm N}$. These results can be better understood
considering the sensitivity of the measurement (actually this is the
square of the sensitivity):
\begin{equation}
{\cal S} =
\frac{\hbox{Var}[\Theta_\alpha]}
{\left( \partial_\phi \langle \Theta_\alpha\rangle \right)^2}
=
\frac{1 - \langle \Theta_\alpha\rangle^2}
{\left( \partial_\phi \langle \Theta_\alpha\rangle \right)^2},
\end{equation}
that is the ratio between the fluctuations of
$\langle \Theta_\alpha\rangle$ and how $\langle \Theta_\alpha\rangle$
varies with respect to $\phi$; $\sqrt{\cal S}$ represents the
smallest change of $\phi$ that can be detected with our measurement
(up to the statistical scaling, of course). In the present case:
\begin{equation}\label{CR:PhEst}
{\cal S}(\phi,\Delta^2)=
\frac{1- e^{-2\Delta^2} \cos^2(\alpha-\phi) \sin^2 2\theta}
{e^{-2\Delta^2} \sin^2(\alpha-\phi) \sin^22\theta},
\end{equation}
which is just the inverse of Eq.~(\ref{fish:qb}): the maximum of
$F$ (maximum information) corresponds to the case of maximum
sensitivity (minimum of $\cal S$). If $\Delta^2=0$ one finds
that $\hbox{Var}[\Theta_\alpha]$ and
$\left( \partial_\phi \langle \Theta_\alpha\rangle \right)^2$ are
always equal, no matter the values of $\alpha$ and $\phi$.
When noise is acting, the maximum of $F$ at $\delta=\pi/2$ corresponds
to the minimum of ${\cal S}(\phi,\Delta^2)$, this fact can be also
understood by geometrical means addressing the special case of $\alpha=0$
($\Theta_0 = \sigma_x$). In this case the result of the measurement
carried out out on the probe is just the projection onto the
$x$-axis: for a fixed change $d\phi$, the change of
$\langle \Theta_0 \rangle$ at $\phi=0$ is smaller than the one at
$\phi = \pi/2$ (see Fig.~\ref{f:sigmaX}).
\par
The Fisher information depends on the actual, unknown value of $\phi$.
However, we can perform an {\em adaptive}, two-step method to achieve
the QFI. During the first step, we use a small amount of data
to obtain a rough estimate $\tilde\phi$ of the phase shift, then,
at the second step, we {\em tune} $\Theta_\alpha$ according to the
transformation $\alpha \to \tilde\phi + \pi/2$: the (eventual) repetition
of these two steps allows to reach the QFI limit. The same result can be
obtained by fixing the measurement at a chosen $\alpha$ and 
tuning the probe state by applying a suitable rotation.
%%%%%%%%
\section{Phase-shift estimation for polarization encoded optical
qubits } \label{s:SimExp}
In our proposal the qubit state corresponds to the polarization degree
of freedom of a coherent state. We refer to $\ket{H}$ and $\ket{V}$ 
respectively as horizontal and vertical polarization.
Initially, we set the
polarization at $\ket{+} = \frac{1}{\sqrt{2}}(\ket{H}+\ket{V})$
(equatorial state) and apply the phase shift with eventual Gaussian
noise (as we have shown in section \ref{s:QPE} this is equivalent to a
non dissipative phase noise).  
Then we measure $\sigma_x$, which
corresponds to inserting a half wave plate (HWP) at $22.5^{\circ}$ 
in front of a polarizing
beam splitter (PBS) and to record the number of counts $n_{+}$ and
$n_{-}$ at the two outputs, which correspond to the outcomes $+1$ and
$-1$, respectively.  Once a data sample $X$ has been acquired, one
needs to give an estimation of the phase shift. Here
we consider two strategies: one based on the inversion of the
probability functions, whose uncertainties come from the error
propagation; the other applies the Bayes theorem to retrieve a prior
distribution of phase shift, knowing this distribution we can calculate
all the desired moments.
%%%%%%%
\subsection{Estimation by inversion}
If we send $M$ copies of our state, then we can write:
\begin{equation}
P_{\Delta^2}(\pm 1 | \phi) = \frac{n_{\pm}}{n_{+} + n_{-}},
\end{equation}
and, by inverting Eq.~(\ref{prob:pm}), after some algebra we obtain the
following expression for the phase shift estimator:
\begin{equation}
\phi_{\rm inv} =
\arccos\left( \frac{n_{+} - n_{-}}{n_{+} + n_{-}}\, e^{\Delta^2} \right).
\end{equation}
The uncertainty in the estimation is thus given by:
\begin{equation}
{\rm Var}[\phi_{\rm inv}] = 
\left( \frac{\partial \phi_{\rm inv}}{\partial n_{+}}\right)^2 
\!\! \sigma^2(n_{+})
+
\left( \frac{\partial \phi_{\rm inv}}{\partial n_{-}}\right)^2 
\!\! \sigma^2(n_{-}),
\end{equation}
where $\sigma^2(n_{\pm})$ are the fluctuations of the numbers of outcomes
$n_{\pm}$. It is worth noting that at each shot, i.e., for each sent copy of
the coherent state, the detected number of photons fluctuates
according to the laws of quantum mechanics.
%%%%%%%
\subsection{Bayesian estimation}
Given the data sample $X=\{x_1,x_2,\ldots,x_N\}$, where
$x_k \in \{-1,+1\}$, $\forall k$, we can define the {\em sample} probability:
\begin{equation}\label{sample:p}
P(X|\phi) = \prod_{k=1}^{N} P_{\Delta^2}(x_k|\phi),
\end{equation}
that is the probability to obtain the whole data sample $X$ given
the unknown phase $\phi$. By means of the Bayes theorem one can write
the {\em a posteriori} probability \cite{berihu:09}:
\begin{equation}\label{posterior:p}
P(\phi|X) = \frac{1}{\cal N}\prod_{k=1}^{N} P_{\Delta^2}(x_k|\phi),
\quad {\cal N} = \int_{\Phi}\!\! d\phi\,P(\phi|X),
\end{equation}
where $\Phi$ is the parameter space. The probability (\ref{posterior:p})
is the probability distribution of $\phi$ {\em given} the data sample $X$.
The Bayesian estimator are thus obtained as:
\begin{align}
\phi_{\rm B} &= \int_{\Phi}\!\! d\phi\,\phi\,P(\phi|X),\\
{\rm Var}[\phi_{\rm B}] &=
\int_{\Phi}\!\! d\phi\,(\phi-\phi_{\rm B})^2\,P(\phi|X).
\end{align}
Bayesian estimators are known to be asymptotically optimal, namely,
they allow one to achieve the Cram\'er-Rao bound as the size of the data
sample increases \cite{Hradil1995,Hradil1996,par:QEQT:08,berihu:09,oliv:09}.
\subsection{Monte Carlo simulated experiments}
\begin{figure}[tb]
\includegraphics[width=0.47\columnwidth]{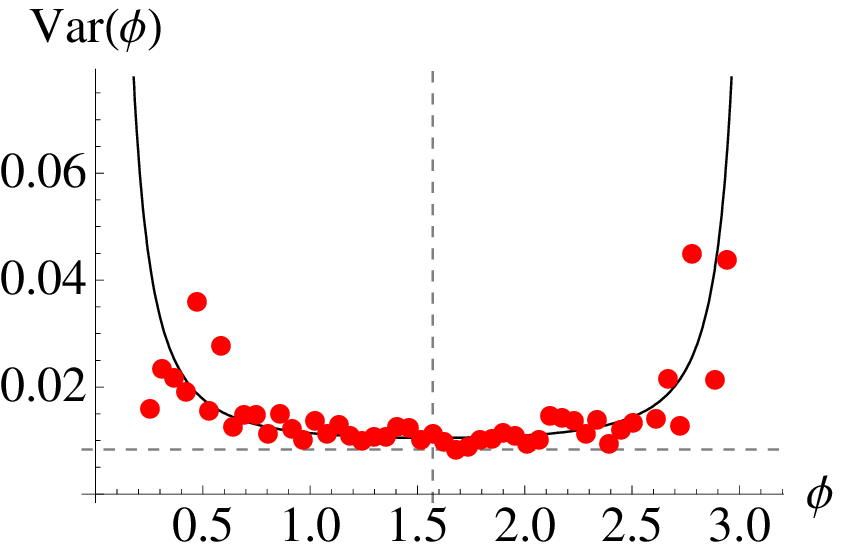}\hfill
\includegraphics[width=0.47\columnwidth]{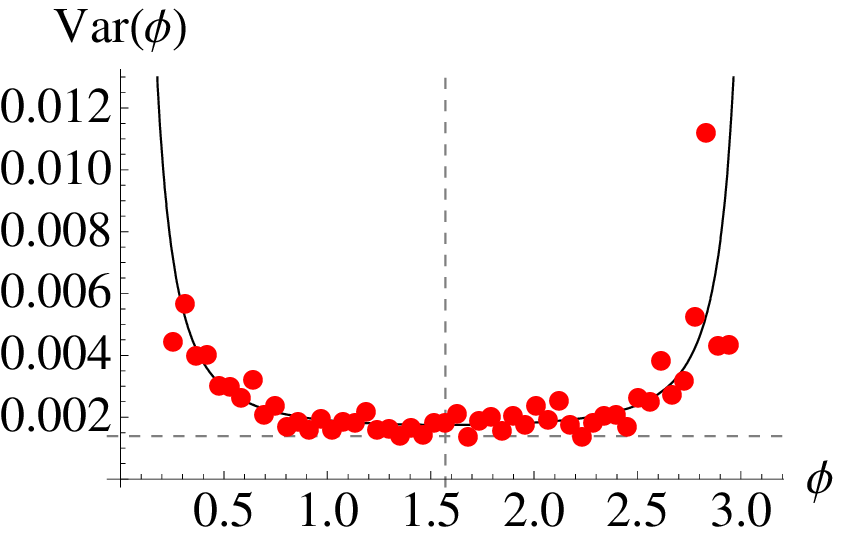}\\
\includegraphics[width=0.47\columnwidth]{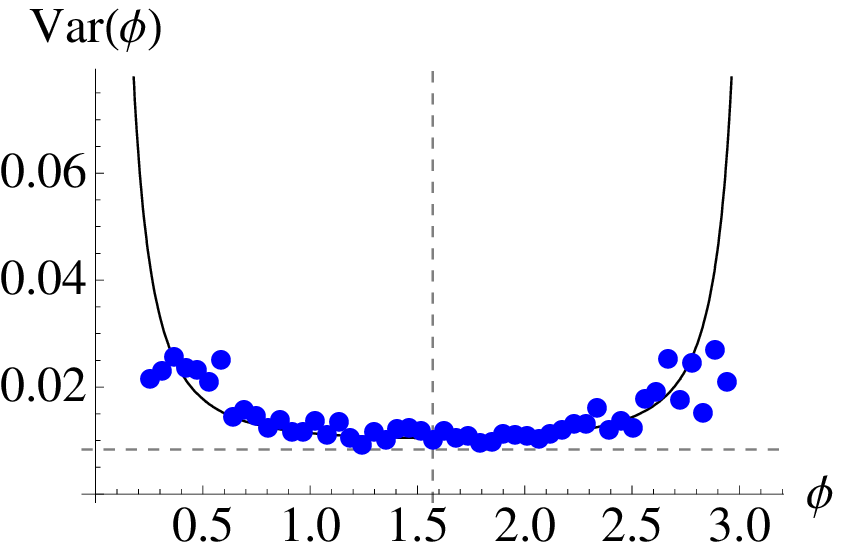}\hfill
\includegraphics[width=0.47\columnwidth]{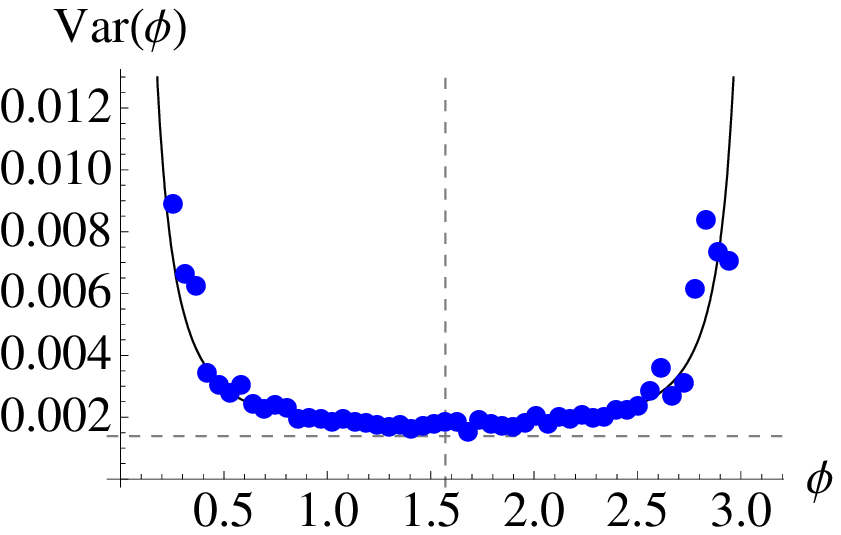}
\caption{\label{f:Sim} Variances from simulated Monte Carlo
experiments performed with $M=60$ copies of the input state and $\Delta=0.34$.
The plots on the top refer to the inversion method, the bottom ones to
Bayesian analysis. For the plots on the left we chose a coherent state 
with an average number of photons equal to $\bar n=2$, for
those on the right we set $\bar n=12$. The solid line is the
Cram\'er-Rao bound $1/(F M \bar n)$. 
The dashed horizontal line is the quantum Cram\'er-Rao bound 
$1/(H M \bar n)$. Notice the
asymptotic optimality of the Bayesian estimation.}
\end{figure}
Fig.~\ref{f:Sim} shows the variances of the estimated phases from Monte
Carlo simulated experiments as function of the unknown phase shift.
The simulations have been performed sending $M=60$ copies of the input
coherent state and choosing $\Delta=0.34$.
The plots on the top refer to the inversion method, the bottom ones to
Bayesian analysis. We investigated two different cases: low energy 
(average number of photons equal to $\bar n=2$, left plots), and high energy 
($\bar n=12$, rights plots).
The solid line is the Cram\'er-Rao bound obtained from Eq.~(\ref{CR:PhEst}), 
while the dashed horizontal line is the 
quantum Cram\'er-Rao bound both rescaled by the effective 
number of measurements $N=M \bar n$.
We can see that in the low energy regime there is a more evident 
deviation from the expected behavior (solid line) at the
extremes of the phase interval $(0,\pi)$. This is the
manifestation of a systematic error due to phase window sampling, 
\emph{i.e.} the tails of the Gaussian distributions, simulating the 
phase-noise, is truncated at the boundary and refolded 
inside the interval.
Notice the asymptotic
optimality of the Bayesian estimation when the energy, and, thus, the
number of events increases. It is also useful to observe how the fluctuations
of the variances obtained by using the Bayesian method are less than the
other ones, which actually depend on the fluctuations of the average
number of photons. A further investigation of Fig.~\ref{f:Sim} leads us to
conclude that both the methods are robust over a large interval of phases:
for the considered examples the variance is almost constant over the
phase interval $(0.5,2.5)$.
%%%%%%%%%%
\section{Experimental setup and results}\label{s:Exp}
\begin{figure}[h!]
\vspace{-0.7cm}
\includegraphics[width=0.75\columnwidth,angle=-90]{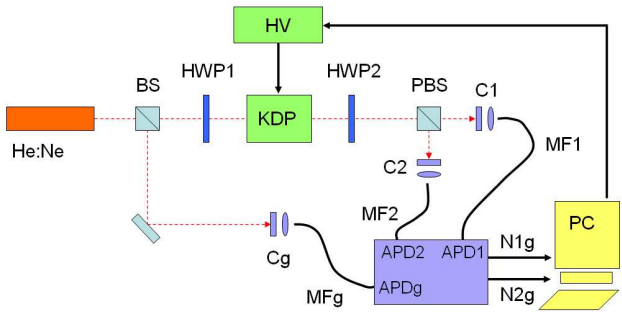}
\vspace{-0.7cm}
\caption{Sketch of the experimental setup to estimate the phase of a
polarization qubit in the presence of non dissipative noise.
The qubit state is set to $\ket{+}$ by inserting a half-wave plate
(HWP1) along the path of the He:Ne laser output, while
a KDP crystal, driven by a high voltage generator (HV),
is used both to add the phase shift $\phi$ and to simulate the noise.
A second half-wave plate @~$22.5^\circ$ (HWP2) together with a
polarizing beam splitter (PBS) implement the measurement (the couplers,
C, and the multi-mode fibers, MF, are used to bring the signals to the
avalanche photodiodes, APD). Finally, the contemporary counts $N_{1g}$
and $N_{2g}$ coming from the APD1 and APD2 and from the APDg, i.e.,
the gate channel signal coming from the beam splitter (BS) placed
in front the laser source, are acquired by a PC module.}
\label{fig:schema}
\end{figure}

The experimental demonstration of our scheme is based on 
a KDP crystal, that allows both the manipulation of the optical 
qubit polarization and the simulation of a phase diffusion environment.  
In Fig.~\ref{fig:schema} we sketch the experimental setup.
\par
A linearly polarized He:Ne laser (Thorlabs HRP120) generates
coherent states that are prepared in the initial
qubit state $|+\rangle = \frac{1}{\sqrt{2}}(|H\rangle+|V\rangle)$
by means of a first half-waveplate (HWP1).
Then the qubits pass through a KDP crystal driven by a stabilized high
voltage generator (HV), up to a maximum of $6$~kV. 
The application of an electrostatic field $E_{z}$ along the $z$ axis of
the KDP (that is also the direction of propagation of the qubit),
a phase shift $\phi(E_{z})$ is introduced between the qubit polarization
components along $x$ and $y$.
The crystal is oriented with the $x$ axis parallel to the horizontal
polarization and tilted around the vertical axis $y$ in order to avoid
the multiple internal reflections superposed with the main beam, that
are removed by a pin hole.
\par
The detection system consists of a second half-waveplate (HWP2),
a polarizing beam splitter (PBS), absorption filters and three detectors.
The HWP2 plate is set at $22.5^\circ$ with respect to the horizontal axis
and the PBS selects the polarization.
Light signals are focused into multi-mode fibers (MF) 
and sent to the detectors APD1 and APD2. The detectors are 
single photon counting modules (SPCM) based on avalanche photodiodes 
operated in Geiger mode with passive quenching. For the coincidence counts 
an electronic circuit based on AND gates is used.
A 50:50 Beam Splitter (BS) is placed before the HWP1 plate to add
a gate channel (g) for the counting measurement. The aim is
to make acquisition with a low number of counts maintaining an
high signal to noise ratio $N_{i}/N_{i,dc}$ ($i=1,2$).
This is the reason for the coincidence counting, according to the
formula ($i=1,2$):
\begin{align}
\label{contemp-counts}
N_{ig} & = (N_{i} + N_{i,dc}) N_{g}\Delta t= N_{i} N_{g}\Delta t +
N_{i,dc} N_{g}\Delta t \nonumber\\  & = N_{ig,true} + N_{ig,dc},
\end{align}
where $N_{ig}$, $N_{i}$ and $N_{g}$ are respectively the coincidence
and direct 
counts on the detectors APD$i$ and APDg, $N_{ig,dc}$ and $N_{i,dc}$ are
respectively the coincidence and direct dark counts, and $\Delta t$ is 
the coincidence time window of the electronic counting module ($\Delta
t = 90$~ns).
$N_{ig}$ could be as low as we want by selecting the time window of the
acquisition and the ratio $N_{ig,true}/N_{ig,dc}=N_{i}/N_{i,dc}$ remain
constant.
We measure about $10^{5}$~counts/s on the detector APDg and a
maximum of about $9\times 10^{4}$~counts/s on the other two but we can
adjust these direct counts changing filters or the detector
voltages above breakdown. The dark counts are below
$200$~counts/s. Each acquisition is taken with a $10$~ms time
window in order to have $\bar n \simeq 10$ in the contemporary
counts. Note that the signal to noise ratio remain $\simeq
10^3$.
The rising and falling time of the high voltage generator are measured around 
$200$~ms so the waiting time between two subsequent measurement is set at
$240$~ms.
\par
We made M=60 acquisitions for each phase-shift $\phi$. Notice that each 
acquisition corresponds to a different random phase
distributed according to a Gaussian distribution centered on
$\phi$ and with standard deviation $\sigma^2(\phi)=\Delta^2/2$
[see Eq.~(\ref{g:noise})]. In our implementation for each
acquisition we send a different voltage to the KDP crystal
according to a proper calibration curve $\phi (V)$.
\begin{figure}[h!]
\begin{tabular}{c}
\includegraphics[width=0.33\textwidth]{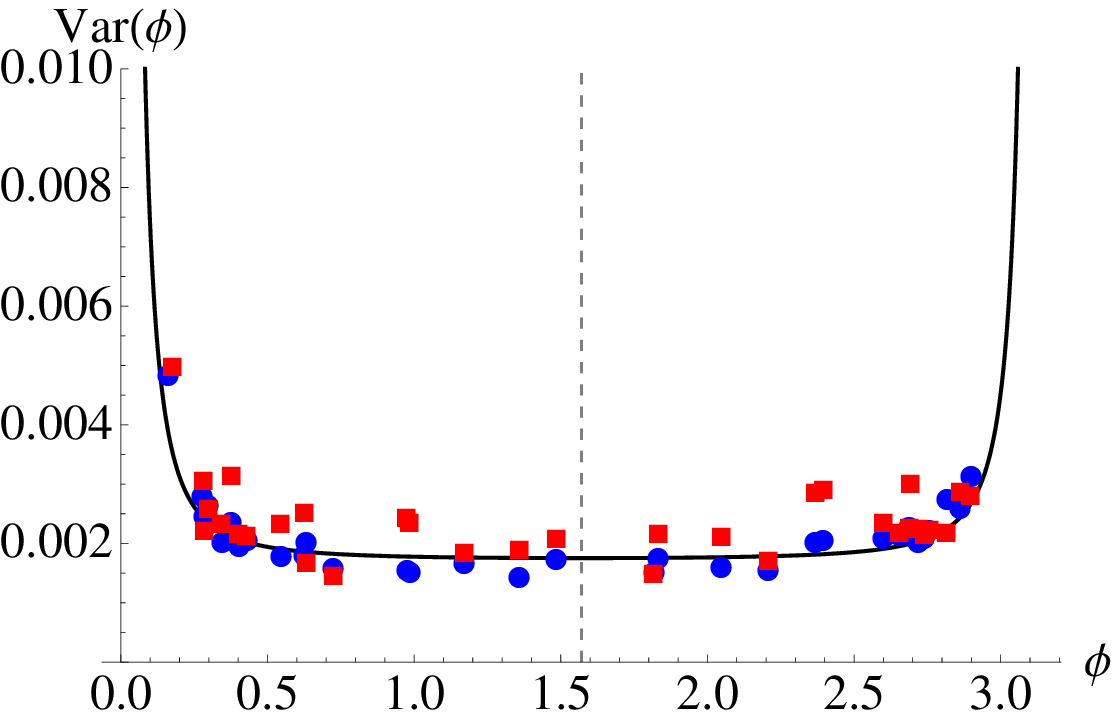}\\
\includegraphics[width=0.33\textwidth]{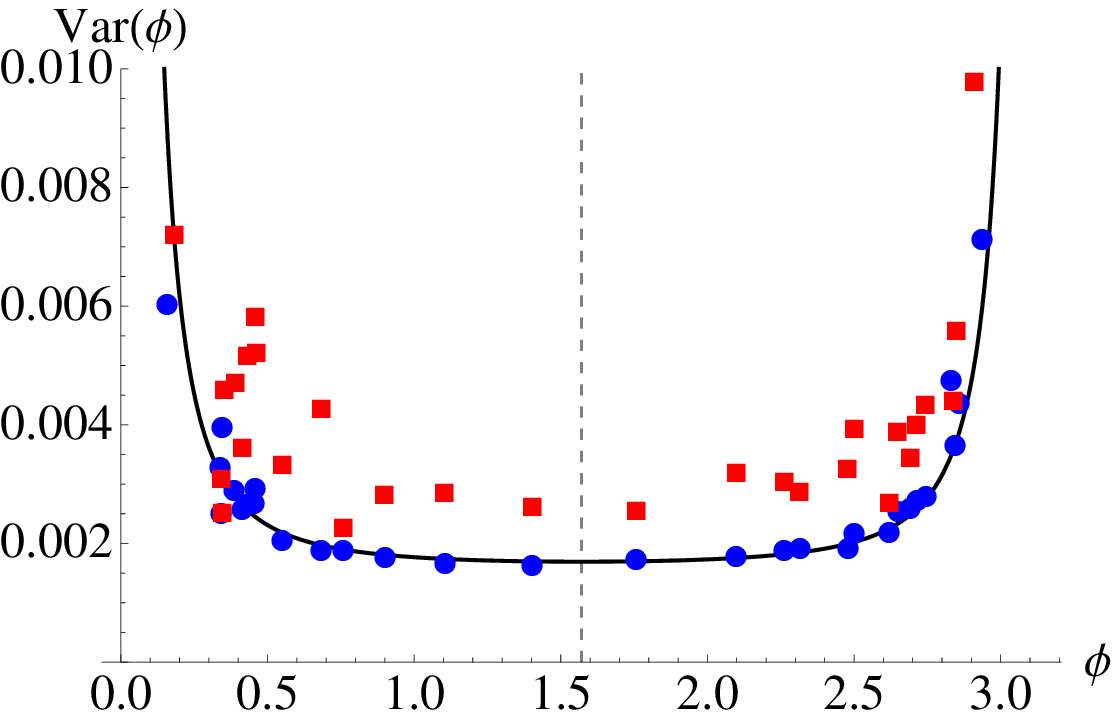}\\
\includegraphics[width=0.33\textwidth]{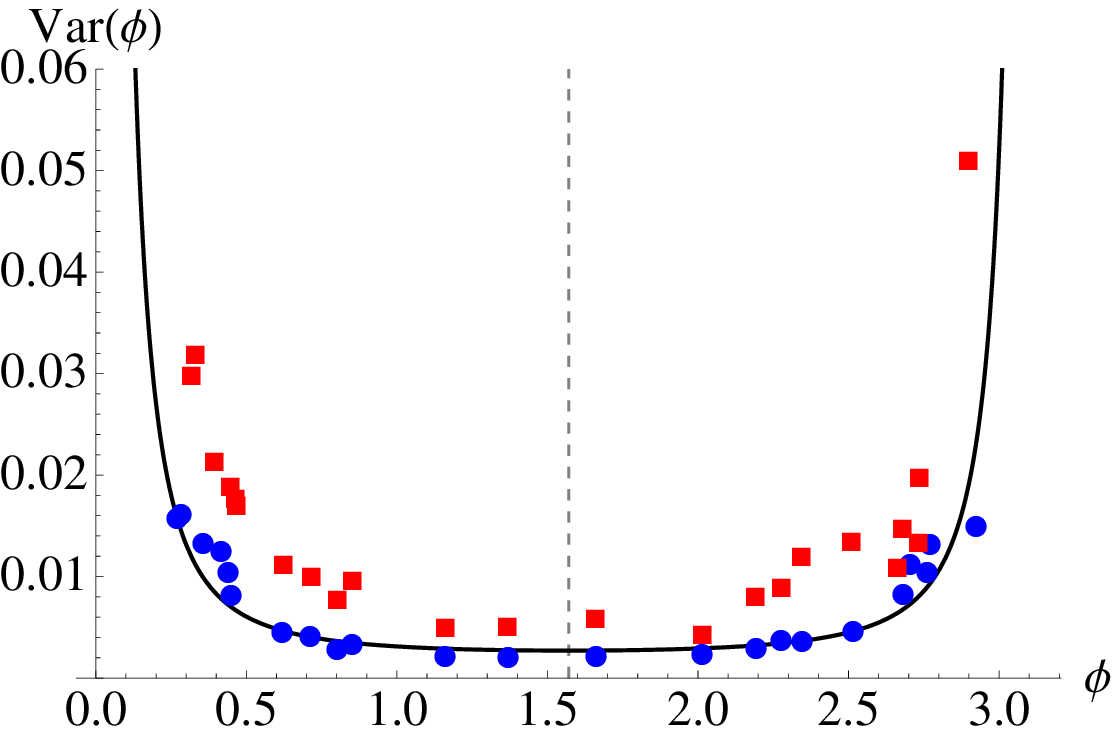}
\end{tabular}
\caption{\label{f:Exp} (Color online) Variances from 
experiments performed with $M=60$ copies of the input state and, from
top to bottom,
$\Delta=0.13\pm 0.02$, $\Delta=0.24\pm 0.03$ and $\Delta=0.48\pm 0.05$.
(Red) Squares refer to inversion method, (blue) circles to Bayesian analysis.
We also have, from top to bottom, $\bar n = 9.83 \pm 0.93$, 
$\bar n = 11.06 \pm 0.43$ and $\bar n = 9.78 \pm 0.95$.
The solid line is the Cram\'er-Rao bound $1/(F M \bar n)$.}
\end{figure} \par
The results are shown in Fig.~\ref{f:Exp}, where we plot the theoretical
Cram\'er-Rao bound (solid line) and the experimental variances obtained
using the inversion method (red squares) and the 
Bayesian analysis ( blue circles).
The Cram\'er-Rao bound is quickly reached by Bayesian
estimation, even for a relatively small number of measurements 
($N=M \bar n$ in the figure, with $M=60$ and $\bar n \simeq 10$).
%%%%%%%%
\subsection{Adaptive method for optimality}
As we have seen Bayesian estimation allows us to reach the Cram\'er-Rao 
bound for any orientation of the measurement angle. On the other
hand, as we pointed out in the section \ref{s:spin}, optimal estimation, 
i.e. with precision at quantum Cram\'er-Rao limit is achieved only 
if $\phi=\pi/ 2$. In this Section we show that ultimate precision
can be always achieved upon the application of an adaptive method 
\cite{oliv:09,mas:00,ha:05}. In the present case,
we proceed as follows: i) we start our estimation obtaining a first
value $\phi_{1}$ of the unknown phase; ii) we set $\alpha$ of the measured
observable (\ref{meas:obs}) to the value $\alpha_1=\pi/2 +\phi_{1}$
(this can be also obtained by adding the phase shift
$\alpha_1'=\pi/2 -\phi_{1}$ to the qubit itself)
so the new phase is near to $\pi/2$. One can also repeat
many times this procedure in order to refine the estimation. However,
as one can see in Figs.~\ref{f:Sim} and \ref{f:Exp}, the optimal 
condition is almost achieved after the first step and the
variance is nearly constant around the optimality region.
We report the results in Table~\ref{t:adaptive}, where one can read
the step number of the adaptive method, the corresponding variance
and the estimated phase. Note that already at the second step optimal 
estimation has been achieved, that is the experimental variances are 
close to the optimal variance given by the quantum Cram\'er-Rao bound.
%%%
\section{Conclusions}
\label{s:concl}
We have demonstrated an optimal estimation scheme for the phase-shift
imposed to a polarization enconded optical qubit in the presence of 
phase diffusion. Our scheme is based on polarization measurement assisted 
by Bayesian estimation and allows to achieve the ultimate quantum bound 
to precision using a limited number of measurements. In turn, Bayes 
estimator is known to be asymptotically unbiased, but for practical 
implementation is of interest to evaluate quantitatively how many 
measurements are needed to achieve the asymptotic region. Our results 
indicate that Bayesian inference represents a useful tool for phase 
estimation.  An adaptive method to achieve optimal estimation in any 
working regime, i.e.  for any value of the unknown phase-shift, has been 
also analyzed in  details and experimentally demonstrated. As a future 
perspective we foresee the possibility to investigate the effects of 
different kind of noises and to employ entangled probes to increases 
the overall stability of the estimation procedure, a results  which 
has been theoretically established for the ideal (i.e without noise) 
case \cite{berihu:09}.
%%%
\begin{table}[h!]
\begin{tabular}{cll}
\toprule
step & ${\rm Var}(\phi)$ & $\phi^{(est)}$ \\
\colrule
$1$ & $4.21\times 10^{-3}$ & $0.33 \pm 0.06$ \\
$2$ & $2.49\times 10^{-3}$ & $0.22 \pm 0.05$ \\
$3$ & $2.52\times 10^{-3}$ & $0.19 \pm 0.05$ \\
$4$ & $2.59\times 10^{-3}$ & $0.20 \pm 0.05$ \\
\botrule
\end{tabular}
\caption{Phase estimation using the adaptive method (see the text
for details). The actual value of the phase is $\phi = 0.17 \pm 0.01$. 
The amplitude of the phase noise is $\Delta = 0.46 \pm 0.06$ and the energy
of the coherent state is $\bar n = 10.97 \pm 0.67$. For each step
we used $M=55$ repetitions for the estimation. For these values of the 
parameters the quantum Cram\'er-Rao bound, evaluated from Eq. 
(\ref{QFI:qubit}) and rescaled by the number of measurments $N=M \bar n$, 
 is equal to ${\rm Var}(\phi) \simeq 2.52 \times 10^{-3}$  }
\label{t:adaptive}
\end{table}
\par
\section*{Acknowledgments}
SO and MGG thanks A. Smerzi and P. Hyllus for discussions.
This work has been partially supported by the CNR-CNISM convention.
%%%%%%%%%%%%%%%%%%%%%%%%%%%%

%%%%%%%%%%%%%%%%%%%%%%%%%%%%

\begin{thebibliography}{50}
\bibitem{op1} A. M. Childs et al., Phys. Rev. A  {\bf 64} 012314 (2001).
\bibitem{op2} M. W. Mitchell et al., Phys. Rev. Lett. {\bf 91} 120402 (2003).
\bibitem{op3} G. M. D'Ariano, P. Lo Presti, Phys. Rev. Lett. {\bf 91}
047902 (2003); G. M. D'Ariano et al., J. Phys. A {\bf 34}, 93 (2001).
\bibitem{op4} J. L O'Brien et al., Phys. Rev. Lett. {\bf 93}, 080502 (2004).
\bibitem{op5} S. H. Myrskog et al., Phys. Rev. A {\bf 72} 013615 (2005).
\bibitem{qtm} M. G. A. Paris and J. Rehacek (Eds), "Quantum State Estimation", 
Lect. Not. Phys. {\bf 649} (2004).
\bibitem{gtm} G. M. D'Ariano et al., Journ. of Phys. A, {\bf 34}, 93 (2001).
\bibitem{mkq} K. Banaszek et al., Phys. Rev. A {\bf 61} 010304(R) (1999).
\bibitem{kw3} D. F. V. James et al., Phys. Rev. A {\bf 64}, 052312 (2001).
\bibitem{hr99} Z. Hradil, Phys. Rev. A {\bf 55} R1561 (1999).
\bibitem{cole05} J. H. Cole et al., Phys. Rev. A {\bf 71} 062312 (2005).
\bibitem{cole06} J. H. Cole et al., Phys. Rev. A {\bf 73} 062333 (2006).
\bibitem{CR:1} H.~P.~Yuen, and M.~Lax, IEEE Trans. Inf. Th. {\bf 19},
740 (1973).
\bibitem{CR:2} C.~W.~Helstrom, and R.~S.~Kennedy, IEEE Trans. Inf. Th.
{\bf 20}, 16 (1974).
\bibitem{CR:3} S.~Braunstein, and C. Caves, Phys. Rev. Lett. {\bf 72},
3439 (1994).
\bibitem{CR:4} S.~Braunstein, C.~Caves, and G.~Milburn, Ann. Phys.
{\bf 247}, 135 (1996).
\bibitem{par:QEQT:08} M.~G.~A.~Paris, Int. J. Quant. Inf. {\bf 7},
125 (2009).
\bibitem{NC} N.~A.~Nielsen and I.~L.~Chuang, {\em Quantum Computation
and Quantum Information} (Cambridge University Press, 2000).
\bibitem{bures:1} D.~J.~C.~Bures, Trans. Am. Math. Phys. {\bf 135}, 199
(1969).
\bibitem{bures:2} A.~Uhlmann, Rep. Math. Phys. {\bf 9}, 273 (1976).
\bibitem{bures:3} M.~H\"ubner, Phys. Lett. A {\bf 163}, 239 (1992).
\bibitem{bures:4} R.~Josza, J. Mod. Opt. {\bf 41}, 2315 (1994).
\bibitem{bures:5} P.~B.~Slater, J. Phys. A {\bf 29}, L271 (1996); Phys.
Lett. A {\bf 244}, 35 (1998).
\bibitem{bures:6} M.~J.~W.~Hall., Phys. Lett. A {\bf 242}, 123 (1998).
\bibitem{bures:7} J.~Dittmann, J. Phys. A {\bf 32}, 2663 (1999).
\bibitem{somm:03} H.-J.~Sommers, et al., J. Phys. A {\bf 36}, 10083 (2003).
\bibitem{yurke86} B. Yurke et al., Phys. Rev. A {\bf 33}, 4033 (1986).
\bibitem{pezze07} L. Pezz\'e et al., Phys. Rev. Lett. {\bf 99} 223602
(2007).
\bibitem{Hradil1995}Z. Hradil, Phys. Rev. A {\bf 51}, 1870 (1995).
\bibitem{Hradil1996}Z. Hradil et al., Phys. Rev. Lett. {\bf 76}, 4295 (1996).
\bibitem{berihu:09} B. Teklu et al., J. Phys. B {\bf 42}, 0335502 (2009).
\bibitem{oliv:09} S.~Olivares and M.~G.~A.~Paris, J. Phys. B {\bf 42}, 055506 (2009).
\bibitem{mas:00} R.~D.~Gill and S.~Massar, Phys. Rev. A {\bf 61},
042312 (2000).
\bibitem{ha:05} M.~Hayashi and K.~Matsumoto, {\em Asymptotic Theory of
Quantum Statistical Inference: Selected Papers}, Ed. M.~Hayashi
(Singapore, World Scientific, 2005) p. 162, chapter 1
(arXiv:quant-ph/0308150).
\end{thebibliography}
\end{document}